\documentclass{elsart}

\usepackage{amssymb}
\usepackage{amsmath}
\usepackage{graphicx}
\usepackage{color}

\newcommand{\be}{\begin{equation}}
\newcommand{\ee}{\end{equation}}
\newcommand{\ben}{\begin{eqnarray}}
\newcommand{\een}{\end{eqnarray}}

\begin{document}

\begin{frontmatter}

\title{Finite-size effects on the phase transition in a four-
and six-fermion interaction model}

\author[UFBA]{L.M. Abreu},
\ead{luciano.abreu@ufba.br}
\author[UVic]{F.C. Khanna}
\ead{khannaf@uvic.ca}
\author[CBPF]{A.P.C. Malbouisson\corauthref{cor2}}
\ead{adolfo@cbpf.br}
\author[UFBA]{J.M.C. Malbouisson},
\ead{jmalboui@ufba.br}
\author[UNB]{A.E. Santana}
\ead{asantana@unb.br}

\corauth[cor2]{Corresponding author}

\address[UFBA]{Instituto de F\'{\i}sica, Universidade Federal da
Bahia, 40210-340, Salvador, BA, Brazil}
\address[UVic]{Department of Physics and Astronomy, University of Victoria,
Victoria, BC, V8P 5C2, Canada and TRIUMF, Vancouver, BC, V6T 2A3,
Canada}
\address[CBPF]{Centro Brasileiro de Pesquisas F\'{\i}sicas, MCTI,
22290-180, Rio de Janeiro, RJ, Brazil}
\address[UNB]{International Center for
Condensed Matter Physics, Instituto de F\'{\i}sica, Universidade de
Bras\'{\i}lia, 70910-900, Bras\'\i lia, DF, Brazil}

\begin{abstract}
We consider four- and six-fermion interacting models at finite
temperature and density. We construct the corresponding free
energies and investigate the appearance of first- and second-order
phase transitions. Finite-size effects on the phase structure are
investigated using methods of quantum field theory on toroidal
topologies.
\end{abstract}

\end{frontmatter}

\section{Introduction}

Temperature and finite-size effects on the phase structure of
physical systems can be described by quantum-field models defined in
spaces with toroidal topologies. The starting point is the quantum
theory of systems at finite temperature, both in the imaginary-time
(Matsubara)~\cite{3mats1,3ume4} and real-time~\cite{3dj,3ume2}
formalisms. Since correlation functions should satisfy periodicity
conditions on the time coordinate, known as Kubo-Martin-Schwinger
(KMS) conditions, the finite-temperature theory is defined on the
compactified manifold $\Gamma _{4}^{1}=\mathbb{S}^{1}\times
\mathbb{R}^{3}$, where $\mathbb{S}^{1}$ is a circumference with
length proportional to the inverse of the temperature and
$\mathbb{R}^{3}$ is the Euclidean 3-dimensional space.
Compactification of spatial dimensions~\cite{birrel1,Ademir} is
considered in a similar way. An unified treatment, generalizing
various approaches dealing with finite-temperature and
spatial-compactification concurrently, has been
constructed~\cite{livro,AOP11,review} by rigourously considering
field theories on toroidal topologies $\Gamma _{D}^{d}=(\mathbb{S}
^{1})^{d}\times \mathbb{R}^{D-d}$, with $d$ ($\leq D$) being the
number of compactified dimensions and $D$ the dimension of the
space-time.

These methods have been employed to investigate spontaneous
symmetry-breaking induced by temperature and/or spatial constraints
in some bosonic and fermionic models describing phase transitions in
condensed-matter, statistical and particle physics; for instance,
for describing the size-dependence of the transition temperature of
superconducting films, wires and grains~\cite{JMP05,linhares}; for
investigating size-effects in first- and second-order
transitions~\cite{Isaque,PRD12,AMM2,Emerson}; and for analyzing size
and magnetic-field effects on the Gross-Neveu (GN)~\cite{gn1} and
the Nambu-Jona-Lasinio (NJL)~\cite{NJL} models, taken as effective
theories~\cite{Weinberg1} for hadronic physics~\cite{AGS,AMM,AMM3}.

In this paper, we consider the massive GN model, modified by the
inclusion of a $\eta({\bar \psi} \psi)^3$ term in the
Hamiltonian~\cite{fi6(1)}, at finite temperature and density. This
model has been considered to investigate color superconductivity,
with possible applications to neutron-star structure~\cite{steiner}
and to chiral symmetry-breaking in hadronic systems~\cite{ruta}.
Also, a rich phase structure, including BCS- and BEC-like phases, in
strongly interacting matter has been obtained with an extended NJL
model having six-fermion interaction~\cite{buballa}. Besides, such
kind of interaction might be of relevance to systems in condensed
matter where GN and NJL models are used, like graphene~\cite{roy}
and quantum phase transitions~\cite{chamati}. In all these cases,
fluctuations due to finite-size play an important role in the phase
structures.

Here, our aim is to discuss the finite-size effects in the extended
GN model by considering compactification of spatial coordinates. We
investigate the existence of first- and second-order phase
transitions in this model and study their dependence on the relevant
parameters: the temperature, chemical potential and the size of the
system. We perform a study of the critical behavior of this model as
a field theory on a toroidal space. We take the space-time dimension
$D=4$ and consider the particular case of two compactified
dimensions ($d=2$), imaginary time  and one spatial coordinate. We
find that, besides the usual symmetry restoration which occurs in
the standard GN model, inverse symmetry-breaking happens in the
extended model; in both cases, there exists a minimal size of the
system below which the phase transition disappears.

\section{The model}

In a $D$-dimensional Euclidian manifold, ${\mathbb R}^{D}$, we
consider the Hamiltonian for a modified massive GN model, which
includes a six-fermion
coupling~\cite{fi6(1),steiner,ruta},
\begin{eqnarray}
H & = &\int d^{D}x\left\{ \psi^{\dagger}(x)(\gamma ^{j}(i\partial
_{j}))\psi (x)
 -m_{0}^{\prime} \psi^{\dagger}(x)\psi (x)\right. \nonumber \\
&& \left. - \frac{\lambda_{0}}{2}\left[
\psi^{\dagger}(x)\psi (x)\right] ^{2}+\frac{\eta_{0}}{3}\left[
\psi^{\dagger}(x)\psi (x)\right] ^{3}\right\},
\label{GN}
\end{eqnarray}
where $m_{0}^{\prime}$, $\lambda_0$ and $\eta_0$ are the free-space
physical mass and coupling constants at zero-temperature and
zero-chemical potential. The $\gamma$-matrices are elements of the
Clifford algebra and we use natural units, $\hbar =c=k_B=1$. The
mass parameter $m_{0}^{\prime}$ may be taken as positive or
negative, i.e. $m_{0}^{\prime} = \pm m_0$ with $m_0>0$, depending on
the phase we choose to start (at zero temperature and chemical
potential) and the nature of the phase transition.

From Eq.~(\ref{GN}), we determine the finite temperature and density
(chemical potential $\mu$) corrections to the mass, considering one
spatial dimension compactified in a circumference of length $L$,
\begin{equation}
m(T,L,\mu) = m_{0}^{\prime}+\Sigma (T,L,\mu), \label{cri1}
\end{equation}
and to the coupling constants,
$\lambda(T,L,\mu)=\lambda_0+\Pi(T,L,\mu)$ and
$\eta(T,L,\mu)=\eta_0+\Xi(T,L,\mu)$). Then, a free-energy density of
the Ginzburg-Landau type is constructed, as was considered in
Ref.~\cite{KL},
\begin{equation}
{\mathcal{F}} = {\mathcal{F}}_0 + A\,\phi ^2(x) + B\,\phi ^4(x) +
C\,\phi ^6(x), \label{july092}
\end{equation}
where  $A = -m$, $B = -\lambda/2$ and $C=\eta/3$. In this formalism,
the quantity $\phi(x)=\sqrt{\left\langle\psi^{\dagger}(x)\psi
(x)\right\rangle}$, where $\left\langle \cdot \right\rangle$ means
thermal average in the grand-canonical ensemble, plays the role of
the order parameter for the transition.

With our sign convention, to have a first-order transition, we must
have $\lambda>0$, $\eta>0$ (to ensure stability of the system), with
the mass $m$ satisfying the condition $ - 16 \eta m = 3 \lambda^2$.
{In addition, for consistency, to have a first-order phase
transition we must require that $m < 0$.} On the other hand, we can
recover a second-order transition for $\eta=0$, but in this case we
must have $\lambda<0$ and the transition is characterized by $m<0$
in the disordered phase and $m>0$ in the ordered phase. We cannot
have simultaneously $B<0$ and $C=0$ in Eq.~(\ref{july092}),
otherwise the stability of the system is lost. It is to be noted
that the choice of $\pm m_0$ for the free-space mass at zero
temperature and chemical potential corresponds to choosing whether
we start with the system in the ordered ($+m_0$) or the disordered
($-m_0$) phase.

We shall consider the simplest approximation where the coupling
constants are taken as fixed, i.e. $\lambda(T,L,\mu)=\lambda_0$ and
$\eta(T,L,\mu)=\eta_0$, with temperature, chemical-potential and
finite-size changes only appearing in the self-energy corrections of
the mass term. Then, the free-energy density becomes
\begin{equation}
{\mathcal{F}} = {\mathcal{F}}_0 - m(T,L,\mu) \phi ^2(x) -
\frac{\lambda_{0}}{2} \,\phi ^4(x) + \frac{\eta_{0}}{3}\,\phi ^6(x).
\label{FreeEnergy}
\end{equation}

Finite temperature and density corrections to the self-energy can be
evaluated by using the Matsubara imaginary-time formalism. The
Cartesian coordinates are specified by
$x=(x_{1}=\tau,x_{2},\mathbf{x})$, where $\mathbf{x}$ is a
$(D-2)$-dimensional vector. The conjugate momentum of $x$ is denoted
by $k=(k_{1},k_{2},\mathbf{k})$, $\mathbf{k}$ being a
$(D-2)$-dimensional vector in momentum space. The KMS conditions,
carrying the anti-periodicity for fermions, imply that the Feynman
rules are modified by the well-known Matsubara
prescription~\cite{livro},
\begin{equation}
\int \frac{dk_{1}}{2\pi }\rightarrow \frac{1}{\beta }
\sum_{n=-\infty }^{+\infty },\;\;\;\;k_{1}\rightarrow
\frac{2\pi}{\beta } (n+\frac{1}{2} - \frac{i \beta \mu}{ 2 \pi})
\equiv \omega _{n}\,,  \label{Matsubara}
\end{equation}
where $\omega_{n}$ are Matsubara frequencies and $\beta = T^{-1}$;
this corresponds to the compactification of the imaginary time in a
circumference of length $\beta$. We shall also investigate
finite-size effects by considering the compactification of one
spatial coordinate ($x_2$) in a length $L$ using the generalized
Matsubara prescription
\begin{equation}
\int \frac{dk_{2}}{2\pi}\rightarrow \frac{1}{L} \sum_{l=-\infty
}^{+\infty },\;\;\;\;k_{2} \rightarrow \frac{2\pi}{L}
(l+\frac{1}{2}) \equiv \omega _{l}\,. \label{MatsubaraL}
\end{equation}
Here we have chosen anti-periodic boundary conditions for the
spatial compactification, instead of the simpler periodic ones,
because they emerge naturally in the generalization of the KMS
conditions satisfied by correlation functions for fermionic fields
in toroidal topologies~\cite{livro,AOP11}.

\begin{figure}[ht]
\begin{center}
\scalebox{0.5}{\includegraphics {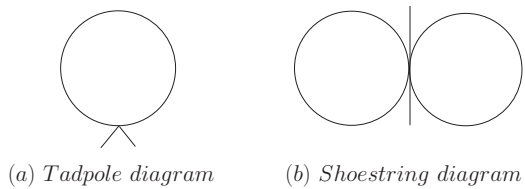} }
\caption{Self-energy corrections to first-order in the coupling
constants $\lambda_0$ and $\eta_0$.} \label{FIG1}
\end{center}
\end{figure}

We consider corrections to the mass to first order in the coupling
constants. This means that there are only two relevant contributions
to the self-energy, one proportional to $\lambda_0$, and other
proportional to $\eta_0$; these contributions correspond the
``tadpole" and the ``shoestring" diagrams shown in Fig.~\ref{FIG1}.
%%%%%%%%%%%%%%
%%%%%%%%%%%%%
Thus, at the lowest order in the coupling constants, the finite
temperature, chemical-potential and size dependent self-energy is
given by
\begin{equation}
\Sigma_{D}(\beta,L,\mu) = \Sigma_{D}^{(a)}(\beta,L,\mu) +
\Sigma_{D}^{(b)}(\beta,L,\mu) ,
\end{equation}
where $\Sigma_{D}^{(a)}(\beta,L,\mu)$ and
$\Sigma_{D}^{(b)}(\beta,L,\mu)$ are respectively the contributions
from the tadpole and the shoestring diagrams. We initially consider
an arbitrary space-dimension $D$ and later reduce to $D=4$.

It should be noticed that the extended GN model is not
perturbatively renormalizable in dimensions greater than two.
However, it can be considered as an effective model and lowest-order
calculations can be performed using an appropriate cut-off or,
equivalently, a convenient minimal subtraction procedure, to discuss
temperature, chemical-potential and size effects on its phase
structure.

\section{$L$-dependent self-energy at finite temperature and density}

We now discuss the effects of finite temperature, density and size
on the mass in order to construct the free-energy for the system.

\subsection{The contribution from the tadpole}

In the Euclidian space, the contribution from the tadpole diagram is
given by
\begin{eqnarray}
\Sigma_D^{(a)}(\beta,L,\mu)\, =\, - \frac{\lambda_{0}}{\beta L}
\sum_{n,l=-\infty }^{\infty } {\rm{Tr} }  \int \frac{d^{D-2}k}{(2\pi
)^{D-2}}\frac{{m_{0}^{\prime}\mathbb{I}_D}}{\mathbf{k}^{2} +
\omega_{n}^{2} + \omega _{l}^{2}+m_{0}^{2}} , \label{sigmaBLa}
\end{eqnarray}
where the symbol $\rm{Tr}$ stands for the trace over spinor space.
The minus sign in the above equation is due to the coefficient of
$\phi^4$ in Eq.~(\ref{FreeEnergy}). Defining the dimensionless
quantities
\begin{equation}
a_{1} = \frac{1}{(m_{0}\beta)^{2}} \; , \;\; c_{1} = \frac12 -
\frac{i\beta\mu}{2\pi} \; , \;\;  a_2 = \frac{1}{(m_0 L)^2} \; ,
\;\; c_2 = \frac{1}{2} \; , \;\; q_{j} = \frac{k_{j}}{2\pi m_{0}} \;
, \label{a1c1qj}
\end{equation}
for $j=3,4,...,D$, we write
\begin{eqnarray}
\Sigma_D^{(a)}(\beta,L,\mu) =  -{\lambda_{0}} \frac{m_{0}^{\prime}
m_0^{D-2}}{4\pi^2} \; {{\rm Tr}\,\mathbb{I}_D}
  \left.
\sqrt{a_{1} a_{2}} \; {\mathcal U}_{2}(s;a_{j},c_{j}) \right|_{s=1}
, \;\; \label{sigmaBLa2}
\end{eqnarray}
where the function ${\mathcal U}_{2}(s;a_{j},c_{j})$ is defined by
\begin{equation}
{\mathcal U}_{2}(s;a_{j},c_{j}) = \sum_{n,l=-\infty }^{\infty } \int
\frac{d^{D-2}q}{\left[ \mathbf{q}^{2} + a_{1} (n + c_{1})^{2} +
a_{2} (l + c_{2})^{2} +b^{2} \right]^s} , \label{U2}
\end{equation}
with $b=(2\pi)^{-1}$.

The integral appearing in the function ${\mathcal
U}_{2}(s;a_{j},c_{j})$ can be treated with dimensional
regularization techniques leading to
\begin{equation}
{\mathcal U}_{2}(s;a_{j},c_{j}) = \pi^{\frac{D-2}{2}}
\frac{\Gamma(\nu)}{\Gamma(s)} \, Y_{2}^{b^2}(\nu;a_{j},c_{j})
,\label{cri32}
\end{equation}
where $\nu = s - (D-2)/2$ and
\begin{equation}
Y_{2}^{b^2}(\nu;a_{j},c_{j}) = \sum_{n_1,n_2= -\infty} ^{+\infty}
\frac{1}{\left[  \sum_{j=1}^{2} a_{j} (n_j + c_{j})^{2} + b^{2}
\right]^{\nu }} \label{epstein2}
\end{equation}
is a double-variable generalized Epstein-zeta function. It is to be
noted that $Y_{2}^{b^2}$ is well-defined only for $\rm{Re} \; \nu >
1$. However it can be analytically continued to the whole complex
$\nu$-plane, becoming a meromorphic function. The analytic
continuation of $Y_{2}^{b^2}$ can be implemented through a
generalized recurrence formula~\cite{ep3,EE}, leading to
\begin{eqnarray}
Y_{2}^{b^2}(\nu;a_{j},c_{j})  =  \frac{\Gamma(\nu - 1)}{\Gamma(\nu)}
\frac{{\pi}}{\sqrt{a_1 a_2}} (b^2)^{1 - \nu}
  + \, \frac{4 \pi^{\nu} b^{1 - \nu}}{\Gamma(\nu) \sqrt{a_1 a_2}}
 \, {\mathcal R}_2(\nu;a_{j},c_{j}) ,
   \label{eps2}
\end{eqnarray}
where the regular part ${\mathcal R}_2$ is given by
\begin{eqnarray}
{\mathcal R}_2(\nu;a_{j},c_{j}) & = & \sum_{j=1}^{2} \sum_{n_j =
1}^{\infty} \cos(2\pi n_j c_j) \left( \frac{n_j}{\sqrt{a_j}}
\right)^{\nu - 1} K_{\nu - 1} \left(\frac{2\pi b n_j}{\sqrt{a_j}}
\right)
\nonumber \\
&& + 2 \sum_{n_1,n_2=1}^{\infty} \cos(2\pi n_1 c_1) \cos(2\pi n_2
c_2)  \left( \sqrt{\frac{l_1^2}{a_1} + \frac{l_2^2}{a_2}}
\right)^{\nu - 1} \nonumber \\
&& \times K_{\nu -1}\left( 2 \pi b \sqrt{\frac{l_1^2}{a_1} +
\frac{l_2^2}{a_2}} \right) , \label{R2}
\end{eqnarray}
with $K_{\sigma}(z)$ being the modified Bessel function of second
kind.

For $s = 1$, we find that the first term in Eq.~(\ref{eps2})
diverges for even dimensions $D \geq 2$ due to the pole of the
factor $\Gamma(\nu -1)$. To obtain a finite tadpole contribution for
the thermal self-energy at finite density with a compactified
spatial coordinate, we perform a minimal subtraction by suppressing
this polar term and obtain, for $D = 4$,
\begin{equation}
\Sigma^{(a)}(\beta,L,\mu) = -\, {\frac{2 \lambda_0 m_0^{\prime}
m_0^2}{\pi^2} } \, {\mathcal W}_2(m_0 \beta,m_0 L,m_0^{-1} \mu) ,
\label{sigmaBLaF}
\end{equation}
where the function ${\mathcal W}_2(x,y,z)$ is given by
\begin{equation}
{\mathcal W}_2(x,y,z) = {\mathcal K}_1(x,z) +  {\mathcal K}_1(y,0) +
{\mathcal K}_2(x,y,z) , \label{W2}
\end{equation}
while the functions ${\mathcal K}_1$ and ${\mathcal K}_2$ are
defined by
\begin{eqnarray}
{\mathcal K}_1(x,z) & = & \frac{1}{x} \sum_{n=1}^{\infty}
\frac{(-1)^n}{n} \cosh{(x z n)} K_1( x n) , \label{K1} \\
{\mathcal K}_2(x,y,z) & = & 2 \sum_{n,l=1}^{\infty}
\frac{(-1)^{n+l}}{\sqrt{x^2 n^2 + y^2 l^2}} \cosh{(x z n)} \,
K_1\left( \sqrt{x^2 n^2 + y^2 l^2} \right) . \;\; \label{K2}
\end{eqnarray}
Notice that the divergent part, which is subtracted, does not depend
on $\beta$, $L$ and $\mu$, and so it does not interfere in the
temperature, chemical-potential and size effects on the system.
Also, it should be noted that the function ${\mathcal K}_1$ is
negative for all values of $x > 0$, tends to 0 as $x \rightarrow
\infty$ and it is quite insensitive to the value of $z$ in the
interval $0 \leq z < 1$, but it is not well defined, i.e. the series
is not convergent, for $z \geq 1$. The behavior of the function
${\mathcal K}_2$ is similar.

\subsection{The shoestring contribution and the self-energy}

Using the dimensionless parameters defined in Eq.~(\ref{a1c1qj}),
the $D$-dimensional contribution of the shoestring diagram for the
finite temperature and chemical-potential self-energy, with a
compactified spatial dimension, can be written as
\begin{eqnarray}
\Sigma_D^{(b)}(\beta,L,\mu) =  {2 \eta_{0} }
\frac{m_0^{2(D-1)}}{16\pi^4} \,  {({\rm Tr}\,\mathbb{I}_D)^2} \left.
{a_{1} a_{2}} \left[ {\mathcal U}_{2}(s;\{a_{j}\},\{c_{j}\})
\right]^2 \right|_{s=1} . \;\; \label{sigmaBLb2}
\end{eqnarray}
It is to be noted that the shoestring diagram involves
$m_{0}^{\prime 2} = m_0^2$ so that this contribution does not depend
on the sign of $m_{0}^{\prime}$. Repeating the steps described above
and taking $D = 4$, we obtain the finite shoestring contribution as
\begin{equation}
\Sigma^{(b)}(\beta,L,\mu) = {\frac{8 \eta_0 m_0^6}{\pi^4} } \left[
{\mathcal W}_2(m_0 \beta,m_0 L,m_0^{-1} \mu) \right]^2 .
\label{sigmaBLbF}
\end{equation}
Thus, the physical self-energy at finite temperature and density
with a compactified spatial dimension, to first order in the
coupling constants, becomes
\begin{eqnarray}
\Sigma(\beta,L,\mu)  =  -\, {\frac{2 \lambda_0 m_0^{\prime}
m_0^2}{\pi^2} } \,
{\mathcal W}_2(m_0 \beta,m_0 L,m_0^{-1} \mu)\nonumber \\
+\,{ \frac{8 \eta_0 m_0^6}{\pi^4} } \left[ {\mathcal W}_2(m_0
\beta,m_0 L,m_0^{-1} \mu) \right]^2 .
 \label{sigmaBLF}
\end{eqnarray}
It should be noted that, in natural units, $\beta$ and $\mu^{-1}$
have dimension of inverse of mass while the coupling constants
$\lambda_0$ and $\eta_0$ have dimensions of $mass^{-2}$ and
$mass^{-5}$, respectively, in $D = 4$. Since the function ${\mathcal
W}_2(m_0 \beta,m_0 L,m_0^{-1} \mu)$ is dimensionless,
$\Sigma(\beta,L,\mu)$ has dimension of mass, as it should.

Adding the contributions of the tadpole and the shoestring diagrams
we obtain the $(T,L,\mu)$-dependent mass, at first-order in the
coupling constants $\lambda_0$ and $\eta_0$; using the reduced
(dimensionless) parameters
\begin{equation}
t = \frac{T}{m_0} = \frac{1}{m_0 \beta} \; , \;\; \chi =
\frac{1}{m_0 L} \, , \;\; \omega = \frac{\mu}{m_0} \; , \;\; \lambda
= \lambda_0 m_0^2 \; , \;\; \eta = \eta_0 m_0^5 \; ,
\end{equation}
the corrected mass can be written as
\begin{equation}
m(T,L,\mu) =  m_{0}^{\prime} - m_{0}^{\prime} \frac{2
\lambda}{\pi^2}  \, {\mathcal W}_2(t^{-1},\chi^{-1},\omega) +\, |
m_{0}^{\prime} | \, \frac{8 \eta}{\pi^4}  \left[ {\mathcal
W}_2(t^{-1},\chi^{-1},\omega) \right]^2 .\label{MassaL}
\end{equation}
It should be noticed that, due to the properties of the functions
${\mathcal K}_1$ and ${\mathcal K}_2$, by taking the limit $L
\rightarrow \infty$, which corresponds to considering the model at
finite temperature and density in the free space, the corrected mass
becomes
\begin{equation}
m(T,\mu) =  m_{0}^{\prime} - m_{0}^{\prime} \frac{2 \lambda}{\pi^2}
\, {\mathcal K}_1(t^{-1},\omega) +\, | m_{0}^{\prime} | \, \frac{8
\eta}{\pi^4}  \left[ {\mathcal K}_1(t^{-1},\omega) \right]^2
.\label{MassaBm}
\end{equation}

\section{Finite-size effects on the phase transitions}

We now discuss the effects of temperature, chemical-potential and
finite-size on the mass of the system. Replacing the corrected mass,
Eq.~(\ref{MassaL}), into Eq.~(\ref{FreeEnergy}), we obtain the
expression of the free energy density which should be analyzed to
investigate the occurrence of phase transitions.

\subsection{Finite-size effects on the second-order phase transition}

\begin{figure}[th]
%\begin{center}
\includegraphics[{height=5.0cm,width=14.0cm}]{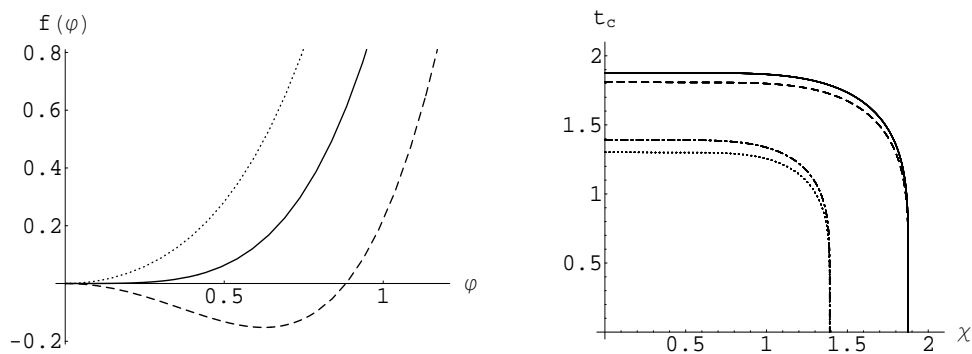}
\caption{{Left panel: free energy density, {$f(\varphi) =
({\mathcal{F}}(\varphi) - {\mathcal{F}}_0)/m_0^4$,} as a function of
the order parameter, {$\varphi = \phi m_{0}^{-3/2}$,} for $\lambda =
2.0$, for the second-order phase transition in free space ($\chi =
0$), with fixed $\omega = 0.0$; dashed, solid and dotted lines
correspond to $t = 1.0,\, 1.876$ and $2.5$, respectively. Right
panel:} reduced critical temperature of the second-order phase
transition as a function of the reduced inverse size for two values
of the chemical potential: $\omega=0.0$ (solid line) and
$\omega=0.9$ (dashed line), for $|\lambda| = 2.0$; and, for the same
values of $\omega$, the dashed-dotted and the dotted lines,
respectively, for $|\lambda| = 4.0$.}
%\end{center}
\label{FIG2}
\end{figure}

In order to have a second-order phase transition in the system with
one compactified dimension, we must take Eqs.~(\ref{FreeEnergy}) and
(\ref{MassaL}) with $\eta_0 = 0$, $\lambda_0 < 0$ and
$m_{0}^{\prime} = + m_0$, that is
\begin{equation}
\frac{m(t,\chi,\omega)}{m_0} = 1 + \frac{2 | \lambda |}{\pi^2} \,
{\mathcal W}_2(t^{-1},\chi^{-1},\omega) \, . \label{cri2X}
\end{equation}
For $\chi = 0$ and $\omega = 0$, corresponding to the model at free
space and zero chemical-potential, it is easy to show that the
coefficient of $\phi^2$ in the free energy density, $-m(t,0,0)$,
increases monotonically from the value $-1$ (for $t=0$), vanishing
at a critical temperature $t_c$, with $m<0$ in the disordered phase
above $t_c$. { In the left panel of Fig.~\ref{FIG2}, the free-energy
density, given by Eq.~(\ref{FreeEnergy}), is plotted as a function
of the order parameter to illustrate the transition. Note that if
one takes $m_{0}^{\prime} = - m_0$, the minimum of the free-energy
density remains zero for all temperatures and no transition occurs.}

For the system with one compactified dimension, from the critical
condition $m(t,\chi,\omega) = 0$, for fixed values of the chemical
potential and of the coupling constant, it follows that the critical
temperature depends on the size of the system; this dependence is {
shown in the right panel} of Fig.~\ref{FIG2}. We find that there is
a minimal size of the system to sustain the broken-symmetry phase,
which is independent of the chemical potential.
%%%%%%%%%%%%%%%%%%
%%%%%%%%%%%%%%%%%%%%%%%%%%%%%%%%
However, this minimal size, $L_0 = (m_0 \chi_0)^{-1}$, depends
strongly on the strength of the quartic self-coupling.

\subsection{Finite-size effects on the first-order phase transition}

Taking the full Eq.~(\ref{MassaL}), with $\lambda_0 > 0$, $\eta_0
> 0$ and fixing {$m_{0}^{\prime} = - m_0$}, there is a
possibility that the system undergoes a
first-order phase transition. In addition, it is required that the
minimum values of the $L$-dependent free-energy density,
Eq.~(\ref{FreeEnergy}), which occur for $\phi$ satisfying $\eta_0
\phi^5 - \lambda_0\phi^3 - m\phi = 0$, should be equal to ${\mathcal
F}_0$, which can be fixed as zero without loss of generality; this
leads to the critical condition
\begin{equation}
-m(T,L,\mu)= \frac{3\lambda_0 ^2}{16\eta_0 } . \label{condicao}
\end{equation}
This critical equation can be written in terms of dimensionless
variables as,
\begin{equation}
{\mathcal G}_2(t,\chi,\omega;\xi) = \frac{\eta}{\lambda^2} ,
\label{critGg2}
\end{equation}
where $\xi = \lambda/\eta$ and the function ${\mathcal G}_2$ defined
by
\begin{eqnarray}
{\mathcal G}_2(t,\chi,\omega;\xi) =  \frac{3}{16} +
\frac{2}{\pi^{2}} \,\frac{1}{\xi} {\mathcal
W}_2(t^{-1},\chi^{-1},\omega)  +\, \frac{8}{\pi^4}
 \left[ \frac{1}{\xi} {\mathcal W}_2(t^{-1},\chi^{-1},\omega)
\right]^2 . \label{FigG2}
\end{eqnarray}

The behavior of the function ${\mathcal G}_2$ is presented in
Fig.~\ref{FigG2}. One can see that the minimum value of ${\mathcal
G}_2$ is $1/16$ while ${\mathcal G}_2(t=0,\chi=0,\omega;\xi) =
3/16$; thus, no solution of the critical equation (\ref{critGg2})
exists if $\eta/\lambda^2 < 1/16${, two solutions occur for $1/16
<\eta/\lambda^2 < 3/16$, while only one solution appears when
$\eta/\lambda^2 > 3/16$}.
%%%%%%
\begin{figure}[th]
%\begin{center}
\includegraphics[{height=7.0cm,width=8.0cm}]{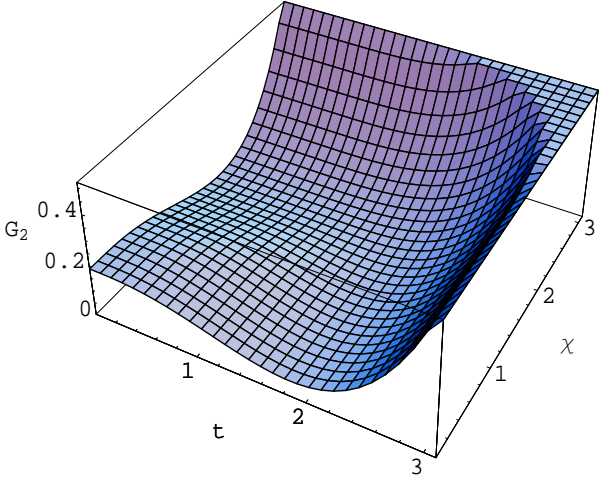}
\caption{Plot of ${\mathcal G}_2$ as a function of $t$ and $\chi$,
for $\omega = 0.5$ and $\xi = 2.0$.}
%\end{center}
\label{FigG2}
\end{figure}
%%%%%%%

\begin{figure}[th]
%\begin{center}
\includegraphics[{height=5.0cm,width=14.0cm}]{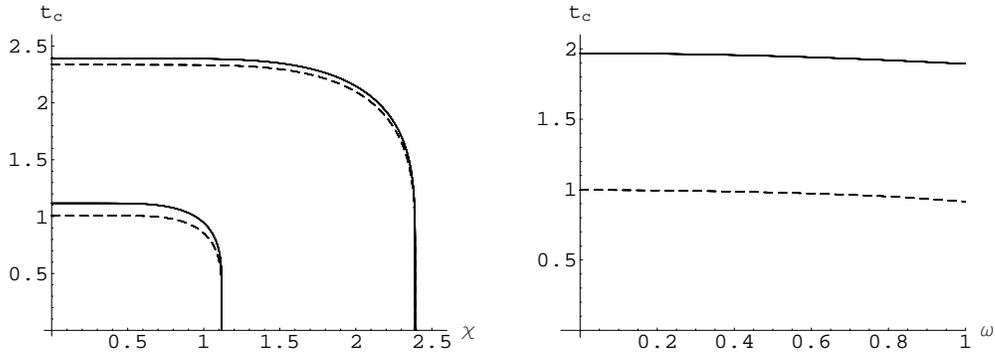}
\caption{{Left panel:} Phase diagrams of first-order
phase-transitions in the $(\chi,t)$-plane, for $\lambda = 4.0$ and
$\eta = 2.0$; the solid and dashed lines correspond to $\omega =
0.0$ and $\omega = 0.9$, respectively. {Right panel: Critical
temperature as a function of the chemical potential for two values
of $\chi$: $0.0$ and $1.96$ (solid and dashed lines, respectively),
taking $\lambda = 4.0$ and $\eta = 4.0$.}}
%\end{center}
\label{Fig4}
\end{figure}

Critical curves in the $(\chi,t)$- {and $(\omega,t)$-planes} can be
obtained simply by determining the level curves of the function
${\mathcal G}_2$, at height $\eta/\lambda^2$, for {fixed values of
the chemical potential and the inverse size}. Phase diagrams for
$\eta/\lambda^2 = 1/8$, corresponding to the case of a
double-transition, are presented in { the left panel of}
Fig.~\ref{Fig4}, taking two values of the chemical potential
$\omega$. {For the single-transition case, the dependence of the
critical temperature on the chemical potential, comparing bulk and
finite-size systems, is shown in the right panel of
Fig.~\ref{Fig4}.}
%%%%%%
%%%%%%%%%%%%%%%%%%

\begin{figure}[th]
%\begin{center}
\includegraphics[{height=4.5cm,width=14.0cm}]{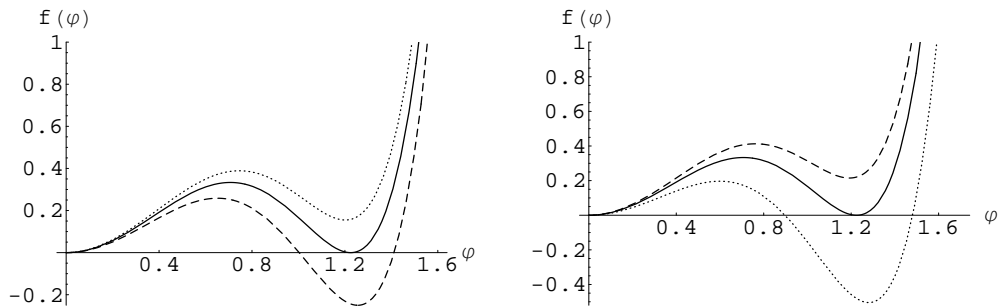}
\caption{The free energy density, {$f(\varphi) =
({\mathcal{F}}(\varphi) - {\mathcal{F}}_0)/m_0^4$,} as a function of
the order parameter, {$\varphi = \phi m_{0}^{-3/2}$,} for $\eta =
2.0$ and $\lambda = 4.0$, for the first-order phase transition in
free space ($\chi = 0$), with fixed $\omega = 0.5$. In the left
panel, dashed, solid and dotted lines correspond to $t = 0.9,\,
1.085$ and $1.2$, and, in the right panel, they correspond to $t =
2.3,\, 2.374$ and $2.5$, respectively. {The left panel corresponds
to a symmetry restoration while the right panel shows an inverse
symmetry breaking.}}
%\end{center}
\label{figFE}
\end{figure}

{For the case of a double-solution (left panel of Fig.~\ref{Fig4}),}
analyzing the free-energy density, we find that, independently of
the value of the chemical potential, the ordered (less symmetric)
phase occurs for low temperatures and large sizes (the inner part of
the smaller curve); in the intermediate region (between the curves),
the stable phase is disordered; while, in the region outside the
larger curve, the stable phase is an ordered one. In other words,
the system reenters the ordered phase at high temperatures and/or
small sizes, corresponding to a kind of inverse symmetry-breaking.
This is illustrated in Fig.~\ref{figFE}, for the system without
spatial compactification.

%%%%%%
%%%%%%%%%%%%%%%%%%

The whole phase structure depends on the values of the coupling
constants but it is also a consequence of the sign chosen for the
physical mass in free space, at zero temperature and chemical
potential. For $\eta/\lambda^2 > 3/16$, there is only one critical
line separating the disordered-phase (low $t,\chi$) from the ordered
one (high $t,\chi$), similar to the larger curves in {the left panel
of} Fig.~\ref{Fig4}{; this is a clear example of inverse-symmetry
breaking. The transition temperature presents a weak dependance on
the chemical potential, as shown in the right panel of
Fig.~\ref{Fig4}; this happens in all situations, for both types of
transition}{, and is a consequence of our approximation of
neglecting corrections of coupling constants}. Also, for all cases
{with $\eta/\lambda^2
> 3/16$ for which} a first-order transition exists in the free
space, there is a minimal length below which the phase-transition
disappears, with the system remaining in the ordered phase {for all
temperatures}.

It should be also pointed out that, as far as estimates of the
transition temperature and the characteristic minimal size are
concerned, these quantities depend on the coupling constants, the
parameters of the model, and on the mass of the fermions, the
natural mass scale of the model. However, independently of the
values of these parameters, an exact relationship can be established
between the free-space transition-temperature ($T_c$) and the
characteristic minimal-size ($L_0$), for vanishing
chemical-potential: due to the symmetry of the function ${\mathcal
W}_2(t^{-1},\chi^{-1},\omega = 0)$ with respect to the variables $t$
and $\chi$, the reduced free-space transition-temperature, $t_c$, is
identical to the reduced inverse minimal-size, $\chi_0$. In any
case, when a first- or a second-order phase transition exists, there
is a characteristic minimal size of the system below which the
transition disappears. Actually, the chemical potential has little
influence on the transition temperature, which decreases slightly as
$\mu$ increases. Then we have in general $t_c \lesssim \chi_0$.
Thus, for both types of phase-transitions and all values of the
chemical potential, we can write, in natural units, $T_c\, L_0
\lesssim 1$.

\section{Conclusions}

We have investigated the appearance of phase transitions in a
modified Gross-Neveu model, which includes a six-fermion
interaction, in the four-dimensional Euclidian space-time. After
constructing  the corresponding free-energy density, we calculate
corrections to the mass term due to finite size, finite temperature
and density.  This leads us to critical equations that yield, for an
appropriate choice of the parameters, to a first- or a second-order
phase transition.

For the standard GN model, not including the six-fermion
interaction, we find a second-order phase transition with the
critical temperature depending strongly on the strength of the
quartic self-interaction. Considering the system with one
compactified spatial dimension, we find that there is a minimal
length below which the phase transition is suppressed and the system
remains in the disordered phase for all values of the temperature;
this characteristic length depends on the strength of the quartic
self-interaction but it is independent of the value of the chemical
potential.

For the extended model, including a six-fermion self-interaction, we
show that a first-order phase transition may exist, separating a
low-temperature condensed phase from a high-temperature disordered
phase, for a range of values of the coupling constants such that
$1/16 < \eta/\lambda^2 < 3/16$. However, in this case, there is also
a second branch of the transition line, as illustrated in {the left
panel of} Fig~\ref{Fig4}, where the system reenters the condensed
phase at even higher temperatures, {a situation corresponding to an
inverse symmetry-breaking}. On the other hand, for $\eta/\lambda^2 >
3/16$ the transition occurs always in this unusual way, with the
low-$T$ phase being disordered and the high-$T$ one being ordered,
while for $\eta/\lambda^2 < 1/16$ no transition exist. In what the
inverse symmetry-breaking is concerned, it is worth to mention that
``exotic" {phase transitions as these we find here, where the high
temperature phase is less symmetric than the low-$T$ one, are known
in the literature. They resemble the inverse symmetry-breaking (or
symmetry nonrestoration) which is found in some scalar multifield
models~\cite{Weinberg74,BL1996,PR2000,Caldas2002,PRP2005,PHAL2007}.
Such unusual behavior has also been found in a gauge theory with an
extra compactified dimension~\cite{ST2009} and for the GN model with
a random chemical potential~\cite{HK2001}.} Further work is needed
to analyze the applicability of the points raised here and to
discuss the effect of corrections to the coupling constants.\\

\begin{center}
{\bf ACKNOWLEDGMENTS}
\end{center}

We thank CAPES, CNPq and FAPERJ, Brazilian Agencies, and NSERC,
Canada.

\end{document}